\documentclass{sig-alternate}
\usepackage{graphicx}
\usepackage[
  pass,
]{geometry}
\makeatletter
\def\@copyrightspace{}
\makeatother
\usepackage{amsmath}
\usepackage{algorithmic}
\usepackage{algorithm}
\usepackage{url}
\usepackage{xspace}
\usepackage{todonotes}
\usepackage{soul}
\usepackage[algo2e,linesnumbered,ruled,vlined,resetcount]{algorithm2e} 

\usepackage{algorithmic}
\usepackage{epstopdf}
\begin{document}
 
\title{Resistance Against Brute-Force Attacks on Stateless Forwarding in Information Centric Networking}
%
%
%
%

\numberofauthors{1} 
%
\author{
%
%
\alignauthor
Bander A. Alzahrani\textsuperscript{{\large \dag}}, Martin J. Reed\textsuperscript{{\large \dag}}, Vassilios G. Vassilakis\textsuperscript{{\large \textasteriskcentered}}\\
       \affaddr{\textsuperscript{{\large \dag}}School of CSEE, University of Essex, Colchester, U.K.}\\
       \affaddr{\textsuperscript{{\large \textasteriskcentered}}Computer Laboratory, University of Cambridge, Cambridge, U.K.}\\
}
\maketitle
\begin{abstract}
Line Speed Publish/Subscribe Inter-networking (LIPSIN) is one of
the proposed forwarding mechanisms in Information Centric Networking
(ICN). It is a stateless source-routing approach based on Bloom
filters. However, it has been shown that LIPSIN is vulnerable to
brute-force attacks which may lead to distributed denial-of-service
(DDoS) attacks and unsolicited messages. In this work, we propose a
new forwarding approach that maintains the advantages of Bloom filter
based forwarding while allowing forwarding nodes to statelessly verify
if packets have been previously authorized, thus preventing attacks on
the forwarding mechanism. Analysis of the probability of attack,
derived analytically, demonstrates that the technique is
highly-resistant to brute-force attacks.
\end{abstract}

\section{Introduction}\label{Introduction}

The Publish-Subscribe Internet Technology architecture (PURSUIT) is
one of the promising ICN candidates for a future Internet. It aims at
redesigning the current Internet to solve many existing limitations
such as security, routing scalability, multicast. The PURSUIT
architecture defines the following three types of network entities:
publishers (\textit{Pub}), subscribers (\textit{Sub}), and mediation
system. The mediation system is broken down into two functions:
\emph{Rendezvous} (\textit{RV})  and \emph{topology management}
(\textit{TM}). These two functions control the third function:
\emph{forwarding} (\textit{FW}). The network connectivity is expressed
by flat, Bloom filter-based identifiers called LIds where each edge in
the network has at least two unidirectional LIds, one in each direction. 


The RV is responsible for matching publishers and subscribers for a
given information item. When a match is detected, the RV contacts the
TM, which is responsible for maintaining intra-domain knowledge of an
autonomous system and to construct a delivery path in the form of a
LIPSIN forwarding identifier (FId) \cite{jokela}. After the path has
been defined, the FW nodes are responsible for packet switching and
delivering the information item from the Pub to the Sub. The POINT
project~\cite{point} builds on this architecture to also introduce a
network attachment point (NAP) for user equipment (UE) to attach to
the network. The UE may be either standard IP clients or
may use LIPSIN for a native ICN interface.  Devices that use native
ICN might include those that are newly developed, for
example Internet of Things (IoT) devices. This paper is relevant to
this latter type of device.

\section{RELATED WORK}\label{Background}
In the LIPSIN forwarding approach  \cite{jokela},  false positives may
exist such that packets can be forwarded over links that were not
intended to be included in the forwarding path; this can be exploited
to launch a \textit{brute-force attack}. In this attack, a malicious
node tries all, or a sufficiently large number of, \textit{possible
  FIds} to obtain one that generates false positives and reaches a
target.  The probability, $p_{fw}$, of guessing a valid FId of a Bloom
filter constructed with a maximum fill factor of $\rho_m$, $k$ hash
functions and representing a path length of $l$ is given by~\cite{jokela}:                                                                       
\begin{equation}\label{pro_of_fp}
p_{fw}=\rho_{m}^{k\cdot l}
\end{equation} 

In \cite{Rothenberg09}, it has been shown that replay attacks and computational attacks are also possible. During a replay attack the attacker exploits a previously created valid FId for sending non-requested traffic. A computational attack is launched by collecting a number of valid FIds and analyzing the correlation between their bit patterns. 

Building upon the LIPSIN forwarding scheme, and prior work \cite{jokela,Rothenberg09}, this paper proposes a forwarding approach that effectively prevents the above mentioned attacks, using network capabilities. In the rest of this paper, we describe our proposed forwarding approach, and analyse the resistance of our solution to brute-force attacks.

\section{SECURE ATTACHMENT APPROACH}
In this approach, we propose a validation mechanism that checks the
legitimacy of FIds sent by a publisher, at the ingress of the
network. The approach is based on the following assumptions: no FW
node in the network is hostile; the FW node that is directly
connected to a user is the NAP; and, each such node holds a pair of 128-bit long master keys, $k_1, k_2$.

\subsection{Secure FId Generation} \label{Secure FId generation}
In the following, we refer to the original forwarding identifier
generated by the TM as FId and its encrypted form is \textit{eFId};
whereas the one used by the Pub is called $eFId_p$ and its decrypted
form is $FId_p$. The hash that is taken over eFId is referred to as
\textit{h}, whereas the hash that is used by the Pub is $h_p$. In the
case of legitimate UE: $eFId_p=eFId$ and $h_p=h$. In this
scheme, the process of generating the FId is almost the same as in
LIPSIN, the only difference is that the constructed FId is sent by the
TM to the NAP instead of the publisher. 
Upon receiving the FId by the NAP, the FId is encrypted using the AES algorithm, which as a result produces an encrypted eFId. The purpose of this encryption step is to preserve the confidentiality of the FId, so that a computational attack is prevented by hiding the content of the FId from the Pub.
 
To prevent brute-force attacks, the NAP node creates a 64-bit hash \textit{h} over the encrypted FId using $k_2$ so that the hash becomes bound to a specific FId. Then, the pair $\{eFId, h\}$ is then forwarded
to the relevant directly connected Pub in order to be used in the
communication with the subscriber. Note that the Pub, which might be a
lightweight device, does not have to compute any encryption algorithms.

\subsection{Secure FId Forwarding}\label{Secure FId forwarding}
Upon receiving the pair $\{eFId, h\}$ from the NAP the publisher
starts the communication with the subscriber by placing this pair in
each transmitted packet header and forwarding it to its local NAP.
When the NAP receives a packet from the Pub, it first performs
two checks: the \textit{security check} and the \textit{forwarding
  check}. The purpose of the security check is to validate the
received $eFId_p$, whether it is legitimate and has been created by
the TM. This check is performed once and only for packets coming from
the publisher. The forwarding check is the LIPSIN membership
check that is performed to decide where packets should be forwarded
for the next hop~\cite{jokela}. An incoming packet is forwarded to the next hop only if it passes these two checks. In the security check, the NAP checks the integrity of the received $eFId_p$. 
  
If the packet passes the security check, then the $eFId_p$ is assumed to be legitimate. In this case the NAP replaces the encrypted $eFId_p$ with a plaintext copy of the $FId_p$. Then, the forwarding check is performed against each outgoing interface using the $FId_p$. If the result of the check is true, then the packet is forwarded to the next FW node along the path. At each subsequent FW node, \textit{only} the forwarding check is performed. To prevent replay attacks, the master key $k_2$ that is used to protect the hash is changed periodically.

\section{ATTACK ANALYSIS} \label{Attack analysis and discussion}
The proposed forwarding approach effectively stops the previously
described attacks. For example, to inject traffic to a victim 4-hops
away  using a brute-force attack, then the attack has to pass both the
security check and the forwarding check at the NAP, and also pass
subsequent forwarding checks in the FW nodes along the path. In this
section, we analyse the probability of injecting unwanted traffic
using brute-force attacks. 
 The probability of passing the forwarding check, $p_{fw}$, is
the probability of guessing a valid FId that causes false positives
along a path, which is given by (\ref{pro_of_fp}). The probability of
passing the security check $p_{sc}$ will now be determined and is
equivalent to guessing the hash using the, so-called, \emph{birthday paradox}
attack \cite{birthday:problem}. To show how a collision is found in
the context of our approach, assume \textit{H} is a hash function such
that $H:D\rightarrow{R}$, where \textit{D} is the set of all possible
combinations of FIds, $R$ is the range of $H$, and $|R|$ = $r$, the
number of all possible hashes. A hash collision occurs when having distinct $eFId_{1}$, $eFId_{2} \in D$  where \textit{H}($FId_{1}$) = \textit{H}($FId_{2}$). To estimate how many attack attempts \textit{x}, consisting of injecting random pairs $\{FId, h\}$, are required to achieve a given probability $p_{sc}$ of finding a hash collision, we use the following approximation \cite{birthday:problem}:

\begin{equation}\label{Ch3Eq:birthday paradox}
x \approx \sqrt{2r \ln\frac{1}{1-p_{sc}}}
\end{equation}

  \begin{figure}[t]
        \centering
\includegraphics[width=.425\textwidth]{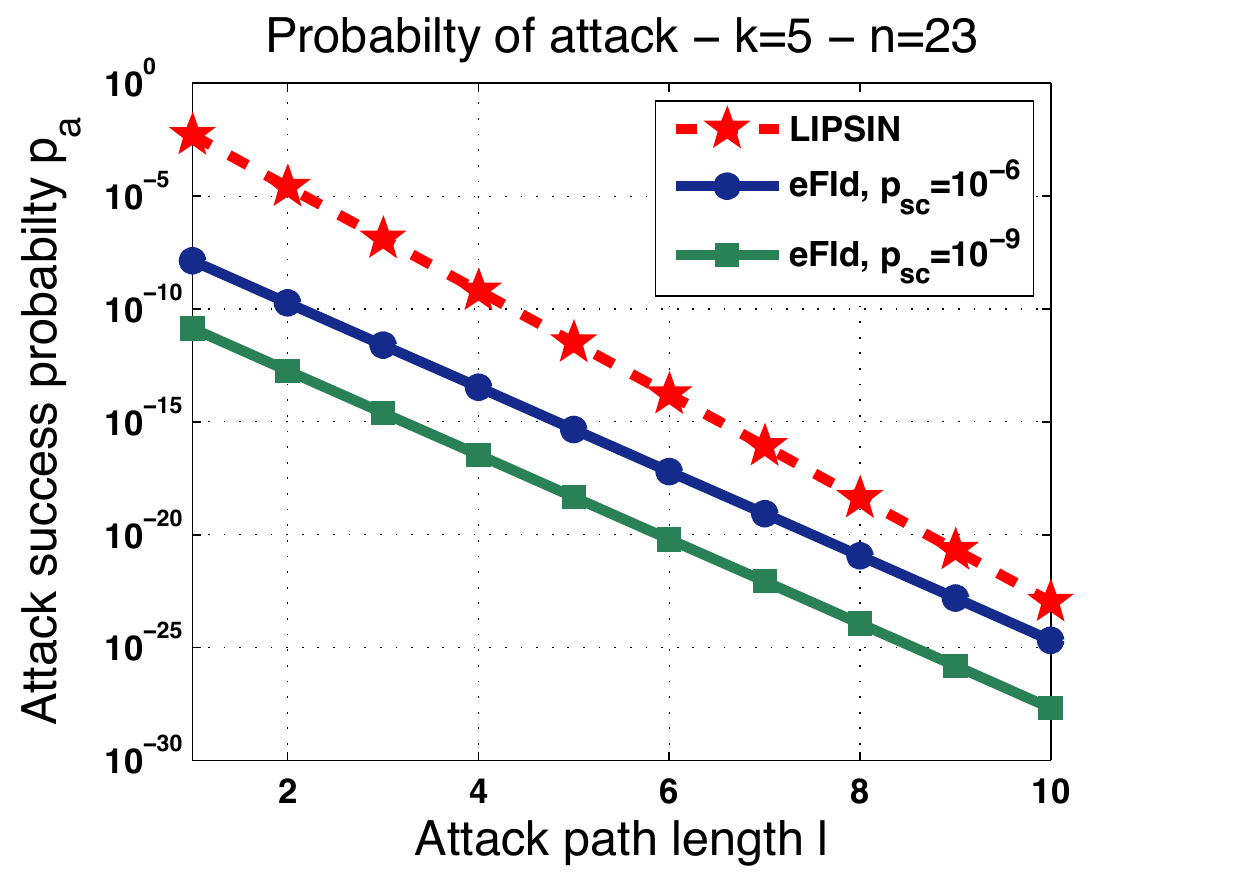} \vspace{-1em} 
\caption{Attack probability, $p_a$, using proposed eFID or LIPSIN
  over different path-lengths.  eFId: $m=256$, $|h|=64$. LIPSIN: $m =
  320$.  Bloom filter has 23 LIds with $k = 5$.}\vspace{-1em}
\label{Ch4fig}
        \end{figure} 
        
Therefore, to successfully reach a victim, the attacker has to pass
all checks at the NAP and the subsequent on-path FW nodes, and
the probability of this is: $p_a=p_{sc} \times p_{fw}$.
          
Figure \ref{Ch4fig} shows the probability $p_a$ for different attack
path lengths $l$ in both approaches: the existing LIPSIN approach and
the proposed NAP approach. The left figure represents the case when
$n=23$ LIds and shows a significant improvement in the probability
$p_a$ when using the encrypted FId. For example, when $p_{sc}$ is $10^{-6}$ the probability of attack $p_a$ to reach a victim attached to the same attacker's NAP node is $\approx1.3 \times 10^{-8}$ compared with approximately $\approx0.0001$ when deploying the basic LIPSIN forwarding approach. This is just to pass the first node on the path, and the probability $p_a$ gets lower as the number of hops increases.

\section{Conclusion}\label{concl}

In this paper, a new approach to protect the forwarding plane against brute-force attacks, computational attacks and replay attacks in the PURSUIT ICN architecture has been presented. This mechanism uses encryption to identify illegitimate forwarding identifiers at the ingress of the network. With this mechanism, the probability of a brute-force attack has been significantly reduced compared to the basic LIPSIN forwarding.


\section{Acknowledgment}
This work has been carried out through the support of the EC project POINT under grant H2020-ICT 643990.


\bibliographystyle{abbrv}
\bibliography{reff}
\end{document}